\documentclass{article}

\usepackage{arxiv}

\usepackage[utf8]{inputenc}
\usepackage[T1]{fontenc}
\usepackage{lmodern}
\usepackage{url}
\usepackage{booktabs}
\usepackage{amsfonts}
\usepackage{amsmath}
\usepackage{amssymb}
\usepackage{nicefrac}
\usepackage{microtype}
\usepackage{graphicx}
\usepackage{listings}
\usepackage{xcolor}
\usepackage[numbers,sort&compress]{natbib}
\usepackage{hyperref}
\usepackage[capitalize,noabbrev]{cleveref}

\hypersetup{
  colorlinks=true,
  linkcolor=blue!60!black,
  citecolor=blue!60!black,
  urlcolor=blue!60!black,
}

\graphicspath{{figures/}}

\lstdefinestyle{codeblock}{
  basicstyle=\ttfamily\footnotesize,
  breaklines=true,
  breakatwhitespace=false,
  columns=fullflexible,
  keepspaces=true,
  showstringspaces=false,
  frame=single,
  framesep=4pt,
  xleftmargin=6pt,
  xrightmargin=6pt,
  aboveskip=6pt,
  belowskip=6pt,
}
\renewcommand{\headeright}{}
\renewcommand{\undertitle}{}

\title{Full-range Binary Classifier Calibration for Stable Model Updates in Production}

\author{%
  Konstantin Berlin \\
  Cisco AI Defense \\
}

\date{}

\begin{document}

\maketitle

\begin{abstract}
  Detection models running in adversarial environments face a malicious distribution that drifts rapidly while the benign distribution stays comparatively stable, so teams retrain and redeploy constantly to stay ahead of new threats.
  Retraining tends to change the output prediction scores, which breaks downstream users of the model.
  For these security-oriented models we need consistent false-positive rate (FPR) across all output values, whereas standard probability-calibration methods target class probability rather than an FPR contract.
  We introduce a method built on top of existing calibration primitives that targets the whole FPR curve, giving scores a consistent FPR meaning across deployments.
  On one held-out split, the observed relative FPR error was at most 2.3\% from 10\% down to 0.1\% FPR and 7.2\% at 0.01\% FPR.
  The shipped artifact remains under 200 KB in measurements across calibration sets from 1K to 10M benign samples.
\end{abstract}

\section{Introduction}
\label{sec:intro}

\begin{figure}[!htb]
  \centering
  \includegraphics[width=0.95\textwidth]{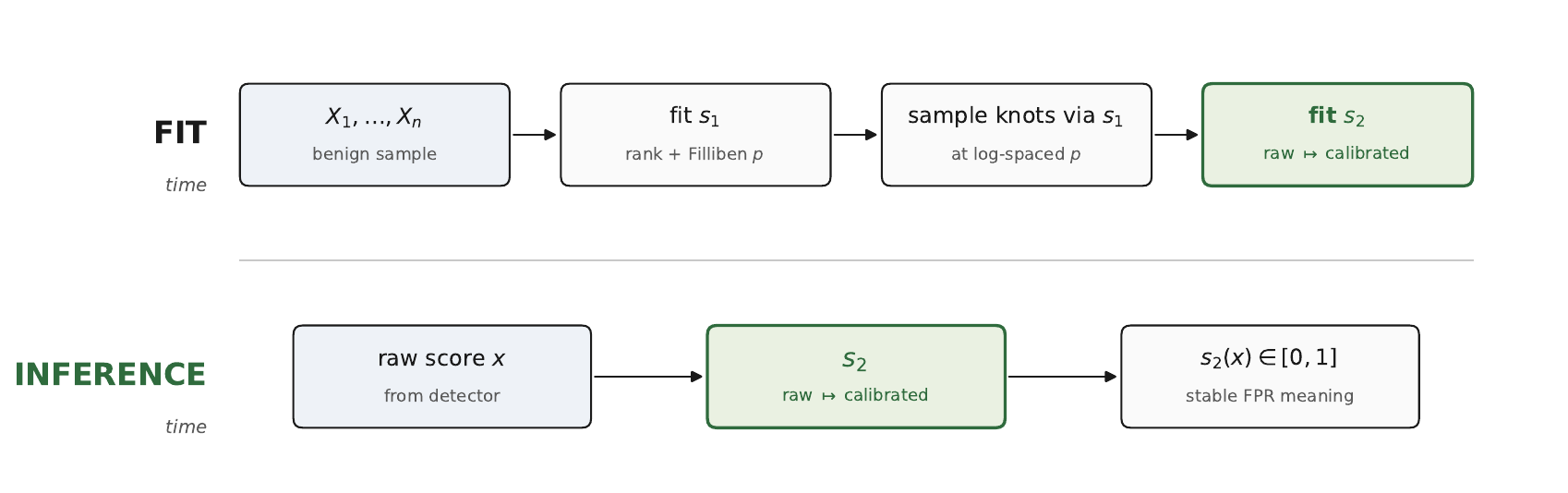}
  \caption{%
    Method overview.
    During fitting, the calibrator sorts benign scores, assigns plotting-position FPR labels, and fits a temporary FPR-to-rescaled-score spline.
    A fixed log-spaced FPR base grid augmented with that spline's fitted edge knots then passes through the spline and the log-scale output contract to produce rescaled-score-to-calibrated-score storage knots.
    The second spline is the shipped sklearn artifact, so inference uses only raw score, fixed scaling, and the rescaled-score-to-calibrated-score map.
  }
  \label{fig:method-overview}
\end{figure}

Detection models running in adversarial environments face a malicious distribution that drifts rapidly while the benign distribution stays comparatively stable.
Teams therefore retrain and redeploy detection models continuously to stay ahead of evolving threats \citep{swanda2025}.
Retraining tends to change the output prediction scores, which breaks downstream users of the model.
For these security-oriented models we need consistent false-positive performance across all output values.
Standard probability-calibration methods such as Platt scaling and isotonic regression calibrate class probability rather than an FPR contract \citep{platt2000,zadrozny2002}.
We introduce a method that calibrates raw scores to the false-positive rate (FPR) directly on benign samples only, composing existing sklearn primitives (\texttt{MinMaxScaler} and \texttt{IsotonicRegression}) into a \texttt{Pipeline} that requires no custom inference code.
The two-spline construction in \cref{fig:method-overview} keeps the FPR-to-rescaled-score lookup at fit time and ships only a raw-score-to-calibrated-score pipeline for production inference.
Here, stability means that each model release is calibrated independently to the same fixed FPR-to-score contract.
Raw-score thresholds may change after retraining, but a calibrated threshold retains the same target benign FPR.

The FPR target depends only on the benign distribution.
In the update setting we target, calibrating to FPR uses only benign traffic, which is the larger and more directly measurable sample pool, and avoids having to characterize adversarial behavior.
In adversarial detection, positive examples are unknown unknowns because new attacks are hard to enumerate and label, yet their count remains far smaller than the benign count.
FPR calibration is robust to positive-count uncertainty because it divides by the large, well-characterized benign count, whereas precision divides by the small, poorly characterized predicted-positive count, as quantified by the binomial planning rule in \cref{sec:reliability}.

A single calibrated score can feed multiple downstream tiers (e.g., block at 0.1\% FPR, alert at 1\%, escalate at 10\%), so calibration must hold across thresholds rather than only at one operating point.

The method's contributions are:

\begin{enumerate}
  \item \textbf{Whole-curve FPR mapping via a non-parametric monotone linear spline.}
    Isotonic regression over the benign empirical CDF maps score thresholds to FPR at every threshold, with no assumed functional form.
  \item \textbf{Fixed log-scale output contract.}
    A piecewise-linear map in $\log_{10}(\mathrm{FPR})$ space pins calibrated score $0.5$ to 0.1\% FPR, $0.7$ to 0.01\%, and $0.85$ to 0.001\%.
    Each step between anchors is a tenfold change in rarity, so the calibrated axis reads uniformly to an operator.
  \item \textbf{Knot-subsampled, fixed-size deployable artifact.}
    A second isotonic fit on a fixed log-spaced FPR base grid augmented with fitted edge knots produces a calibration artifact under 200 KB in our measurements, independent of calibration-set size.
  \item \textbf{Below-floor extrapolation with safe clipping.}
    Linear extrapolation from the two lowest observed $(\mathrm{FPR}, \mathrm{score})$ points, composed through the log anchors, keeps the output monotone and bounded below the sample-supported FPR floor.
  \item \textbf{Release-specific calibration under a fixed cross-retraining contract.}
    Each model release refits its calibration artifact on that model's benign scores while retaining the fixed FPR-to-score anchors, so downstream systems can keep the same calibrated thresholds across releases.
\end{enumerate}

\Cref{sec:evaluation} reports anchor-level calibration error and full-curve diagnostics on the held-out-from-fit subset, with the calibration-fit subset shown separately.

\section{FPR estimate bounds and biases}
\label{sec:reliability}

Two finite-sample effects bound FPR estimation accuracy for any calibrator: sampling variance on the benign set and edge-of-sample bias when rank-selected score thresholds receive FPR labels for spline knots.

\subsection{Sampling variance}
\label{sec:precision}

Let $S_i$ be the model's scalar prediction score on benign calibration input $i$ for $i=1,\ldots,n$, and let $S_{\mathrm{benign}}$ be the score the same model would assign to a fresh benign input from the same distribution.
For a fixed score threshold $\tau$, each benign calibration example either fires or does not fire, so the empirical FPR is a proportion over $n$ independent 0/1 outcomes:
\begin{equation}
\label{eq:fixed-threshold-binomial}
\widehat q(\tau)=\frac{1}{n}\sum_{i=1}^{n}\mathbf{1}\{S_i\ge \tau\}.
\end{equation}
The standard normal approximation for a binomial proportion gives the 95\% interval \citep{nistBinomialConfidence}
\begin{equation}
\label{eq:normal-binomial-interval}
\widehat q(\tau)
\pm
2\sqrt{\frac{\widehat q(\tau)(1-\widehat q(\tau))}{n}}.
\end{equation}
For target FPR $p$ and relative half-width $r$, converting \cref{eq:normal-binomial-interval} gives the planning rule derived in \cref{sec:sample-size-rule-derivation}:
\begin{equation}
\label{eq:relative-sample-size-rule}
n \approx \frac{4}{r^2p}.
\end{equation}
\Cref{tab:sample-sizes} reports \cref{eq:relative-sample-size-rule} for common deployment FPRs.
At a target FPR of 0.1\%, the normal approximation recommends 64{,}000 benign samples, corresponding to a rough 95\% range of 0.075\%--0.125\%.

\begin{table}[!htb]
  \centering
  \caption{%
    Benign sample sizes estimated using the normal approximation.
  }
  \label{tab:sample-sizes}
  \begin{tabular}{rrrr}
    \toprule
    Target FPR $p$ & $r = 50\%$ & $r = 25\%$ & $r = 10\%$ \\
    \midrule
    $10^{-1}$ & 160       & 640       & 4{,}000 \\
    $10^{-2}$ & 1{,}600   & 6{,}400   & 40{,}000 \\
    $10^{-3}$ & 16{,}000  & 64{,}000  & 400{,}000 \\
    $10^{-4}$ & 160{,}000 & 640{,}000 & 4.0M \\
    $10^{-5}$ & 1.6M      & 6.4M      & 40M \\
    $10^{-6}$ & 16M       & 64M       & 400M \\
    \bottomrule
  \end{tabular}
\end{table}

\subsection{Edge-of-sample bias}
\label{sec:bias}

At a fixed score threshold, \cref{sec:precision} measures counting error.
The first calibration spline has a different problem because it starts from sorted benign scores and must attach one FPR label to each rank-selected score threshold.
For a fixed threshold $\tau$, the fresh-sample FPR is
\begin{equation}
\label{eq:true-deployment-fpr}
q(\tau)=\Pr(S_{\mathrm{benign}}\ge\tau),
\end{equation}
where $S_{\mathrm{benign}}$ is the model score of a fresh benign deployment example.

Let $k\in\{1,\ldots,n\}$ count ranks from the largest benign model score downward, and let $\tau_k$ be the $k$-th largest benign model score.
The fresh-sample FPR at that rank-selected threshold is
\begin{equation}
\label{eq:rank-selected-fpr}
q_k=q(\tau_k)=\Pr(S_{\mathrm{benign}}\ge\tau_k).
\end{equation}
The spline label attached to $\tau_k$ is written $\widetilde q_k$ to distinguish it from the empirical FPR estimate $\widehat q(\tau)$ in \cref{eq:fixed-threshold-binomial}.
The naive spline label is
\begin{equation}
\label{eq:uncorrected-rank-position}
\widetilde q_k^{\mathrm{sample}}=\frac{k}{n}.
\end{equation}
\Cref{eq:uncorrected-rank-position} is an in-sample count: $k$ of the $n$ calibration scores are at or above $\tau_k$.
The spline needs the fresh-sample value $q_k$, the probability that a fresh benign score exceeds the threshold selected by that rank.
Because $\tau_k$ was chosen after sorting, it already had to beat $n-k$ calibration scores, so a fresh benign score exceeds it less often than $k/n$ suggests.
With $n$ sorted samples creating $n+1$ probability gaps, the $k$-th largest score has mean fresh-sample FPR \citep{nistOrderStatisticMeans}
\begin{equation}
\label{eq:mean-rank-position}
\widetilde q_k^{\mathrm{mean}}=\frac{k}{n+1}.
\end{equation}
Since $k/(n+1)<k/n$, the naive knot assigns too large an FPR label to $\tau_k$ and returns a threshold that is slightly more selective than requested.

The mean correction shows that the naive label $k/n$ is biased high for the fresh-sample FPR.
The implementation uses Filliben because the spline must choose one typical label per rank, and Filliben is the standard median plotting-position rule.
Filliben uses order-statistic medians rather than means for probability-plot locations \citep{filliben1975}, and the NIST probability-plot formula gives the interior uniform median approximation \citep{nistProbabilityPlot}.
After flipping from lower-tail percentile to upper-tail FPR, the interior-rank label is
\begin{equation}
\label{eq:filliben-position}
\widetilde q_k^{\mathrm{Filliben}}=\frac{k-0.3175}{n+0.365}.
\end{equation}
The implementation uses Filliben by default, with endpoint values $1-0.5^{1/n}$ for the largest benign score and $0.5^{1/n}$ for the smallest.
Held-out evaluation still reports count-based FPR; Filliben is used only to place first-spline knots during fitting.

Filliben reduces relative label bias, not relative sampling noise.
At $p=10^{-3}$ with $50{,}000$ benign calibration samples, the relevant tail rank is about $50$.
Filliben moves the naive label by about $0.6\%$ relative error, while the sampling interval from \cref{eq:normal-binomial-interval} is about $\pm28\%$ relative error.
The correction is justified, but the practical error is dominated by having only about 50 benign tail examples.
\Cref{fig:plotting-position-error} shows that Filliben removes the median relative label error, and \cref{fig:plotting-position-absolute-error} shows that mean absolute relative label error barely changes because rank-selected thresholds still move across calibration sets.
Only more benign calibration samples materially reduce that dominant relative error, as \cref{tab:sample-sizes} shows.

\section{Method}
\label{sec:method}

The calibrator builds a shipped raw-score-to-calibrated-score pipeline from a temporary FPR-to-rescaled-score spline and a fixed FPR-to-calibrated-score contract.
\Cref{fig:method-overview} shows the fit-time objects and the shipped object.
We use \texttt{IsotonicRegression} because it is sklearn's built-in monotone piecewise-linear spline and fits directly inside an sklearn \texttt{Pipeline}.
In this method, it serves mainly as a deployable spline container.
The implementation will be released at \url{https://github.com/cisco-ai-defense/fpr-model-calibration}.

\textbf{Log-scale output anchors.}
Log spacing matches deployment practice because operators perceive FPR changes by order of magnitude rather than by linear differences in probability.
The calibrated axis therefore starts from a fixed contract in $\log_{10}(\mathrm{FPR})$ space before any detector-specific spline is fit.
\Cref{tab:anchors} defines the full anchor map, including the convention that a calibrated score of $0.5$ corresponds to 0.1\% FPR.
Pinning these values gives the calibrated score the same alert-rate meaning across model versions and detector categories.
The pipeline emits calibrated scores no greater than $0.99$, so a threshold of $1.0$ flags nothing.

\begin{table}[!htb]
  \centering
  \caption{Log-scale anchor map from FPR to calibrated score.}
  \label{tab:anchors}
  \begin{tabular}{rrl}
    \toprule
    FPR & Calibrated threshold & Interpretation \\
    \midrule
    100\%    & 0.00 & Flag everything \\
    10\%     & 0.10 & 1 in 10 benign flagged \\
    1\%      & 0.30 & 1 in 100 \\
    0.1\%    & 0.50 & 1 in 1{,}000 \\
    0.01\%   & 0.70 & 1 in 10{,}000 \\
    0.001\%  & 0.85 & 1 in 100{,}000 \\
    0.0001\% & 0.95 & 1 in 1{,}000{,}000 \\
    0\%      & 1.00 & Flag nothing \\
    \bottomrule
  \end{tabular}
\end{table}

\textbf{Scale scores.}
A \texttt{MinMaxScaler} with \texttt{feature\_range=(0, 0.99)} fit on the input domain $[0, 1]$ maps raw scores to $[0, 0.99]$.
Using the fixed input domain keeps production scores above the calibration-time maximum inside the trained range.

\textbf{Spline 1: FPR to rescaled score.}
The first spline answers the fit-time lookup from target FPR $p$ to the rescaled score threshold that should carry it.
The finite benign sample supplies thresholds only at its rank labels, so most desired FPR values on the log grid have no exact observed score.
This temporary spline interpolates those missing thresholds during fitting and is not shipped.
Sort the $n$ rescaled benign scores as $s_{(1)}\le\cdots\le s_{(n)}$.
For score $s_{(j)}$, let $k=n-j+1$ be its rank from the largest score downward.
Attach the Filliben FPR label $\widetilde q_k^{\mathrm{Filliben}}$ from \cref{eq:filliben-position} to $s_{(j)}$.
Fit \texttt{IsotonicRegression(increasing=False)} on these $(\mathrm{FPR\ label}, \mathrm{rescaled\ score})$ pairs.

\textbf{Synthetic knot grid.}
Start from a fixed log-spaced base grid of $\sim 10{,}000$ FPR values, with explicit grid points at every decade from $10^{-10}$ to $1$.
Augment the base grid with Spline 1's high-FPR endpoint and the two lowest-FPR knots that define extrapolation.
For each resulting FPR, query the FPR-to-rescaled-score spline to get a rescaled-score threshold and query the fixed log-scale FPR-to-calibrated-score contract to get the calibrated score.
For FPRs below Spline 1's smallest observed FPR label, replace isotonic clipping with linear extrapolation from the two lowest observed $(\mathrm{FPR}, \mathrm{rescaled\ score})$ knots, bounded inside $[0, 0.99]$.
These generated $(\mathrm{rescaled\ score}, \mathrm{calibrated\ score})$ pairs are storage knots for the shipped spline, not new statistical evidence below the sample-size floor.

\textbf{Spline 2: rescaled score to calibrated score.}
Fit \texttt{IsotonicRegression(increasing=True)} on the generated $(\mathrm{rescaled\ score}, \mathrm{calibrated\ score})$ pairs.
Include boundary training pairs $(0.99,0.99)$ and $(1.0,1.0)$ in the rescaled-score-to-calibrated-score fit.
If tied benign scores produce repeated score knots, \texttt{IsotonicRegression} averages the tied block while preserving monotonicity.
The shipped artifact is an sklearn \texttt{Pipeline} with the \texttt{MinMaxScaler} followed by the second \texttt{IsotonicRegression}, so production inference uses standard sklearn objects rather than custom spline code.

\section{Fixed-size deployable artifact}
\label{sec:artifact}

Production deployment needs artifact storage bounded independently of calibration-set size.
Let $K$ denote the number of breakpoints retained by the fitted \texttt{IsotonicRegression}.
An uncapped fit stores up to one breakpoint per distinct training score, so $K$ can approach $n$.
Its two \texttt{float64} breakpoint arrays require $16K$ bytes; at $K=10$ million, those arrays alone require 160 MB.

The knot-subsampling step in \cref{sec:method} caps $K$ at approximately 10{,}000 regardless of $n$.
The base FPR grid is fixed and log-spaced, not Monte Carlo sampled.
With \texttt{n\_knots=10{,}000}, it places 999 points in every FPR decade from $10^{-10}$ to $1$.
The fit augments that base grid with Spline 1's high-FPR endpoint and two lowest-FPR knots, preserving the fitted boundaries and the points that define low-FPR extrapolation.
These edge locations depend on the fitted sample, whereas the base allocation is fixed.
Log spacing gives every order of magnitude comparable resolution and densely samples the low-FPR tail.
Fixed base placement avoids random gaps in that allocation.

With \texttt{n\_knots=10{,}000}, the production pipeline serialized under \texttt{joblib.dump} to 54--161 KB across calibration sets from 1K to 10M benign samples, remaining below 200 KB.
Dependencies are \texttt{MinMaxScaler}, \texttt{IsotonicRegression}, and \texttt{Pipeline}, with no custom inference code: \texttt{joblib.load(path)} then \texttt{pipeline.predict(scores)}.

\section{Evaluation on Credit Card Fraud Detection}
\label{sec:evaluation}

\begin{figure}[!htb]
  \centering
  \includegraphics[width=0.95\textwidth]{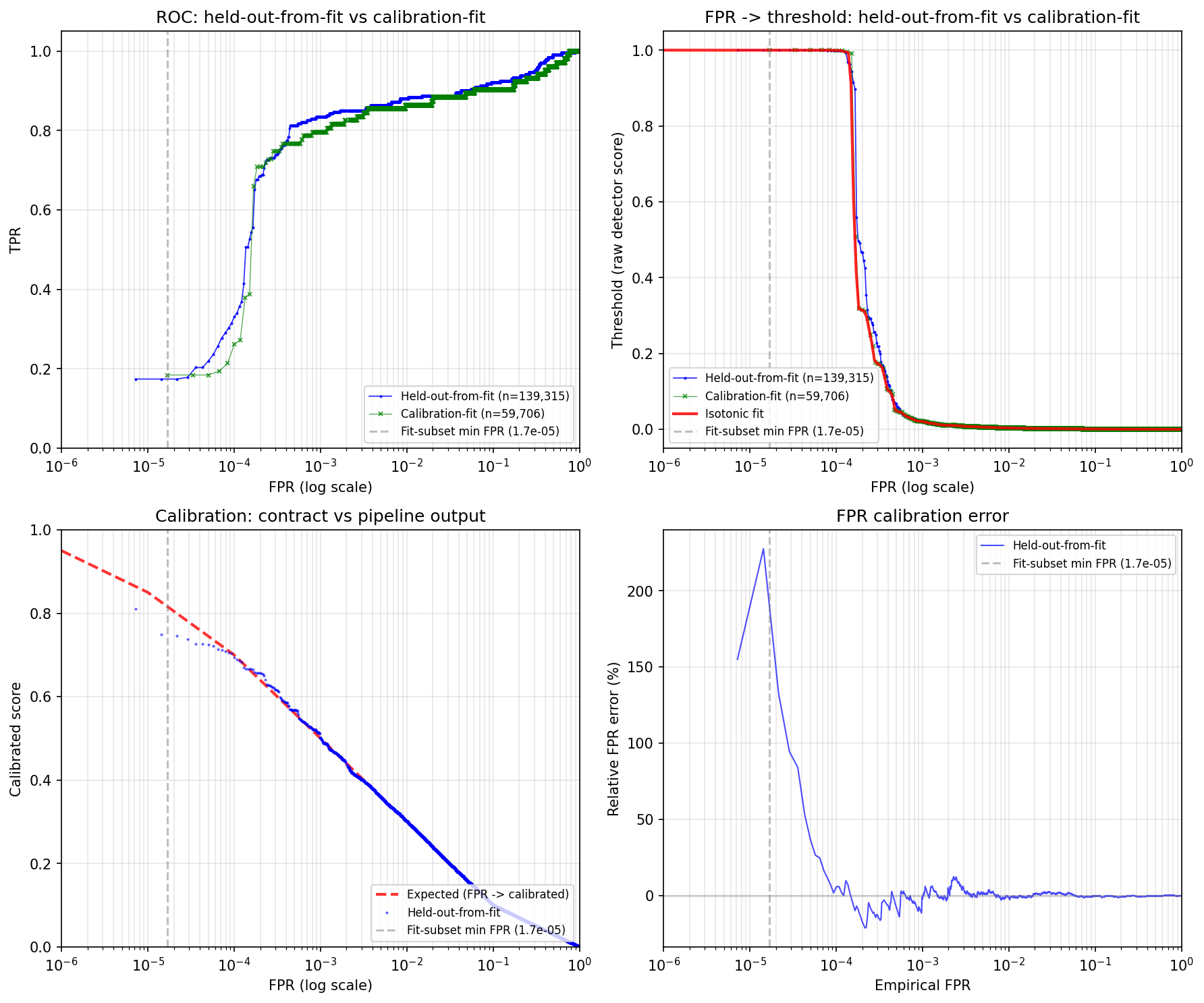}
  \caption{%
    Four-panel diagnostics for the calibration pipeline on the Credit Card Fraud holdout.
    Top-left: empirical ROC (log-FPR axis) on the held-out-from-fit subset in blue and on the calibration-fit subset in green (marked with $\times$), with the fit-subset's minimum-resolvable FPR drawn as a vertical dashed line across every panel.
    Top-right: the empirical FPR-to-threshold relation on the same two populations, with the red line showing the fitted FPR-to-raw-score-threshold spline.
    Bottom-left: expected log-scale contract (red dashed) versus pipeline output on the held-out-from-fit subset (blue).
    Bottom-right: relative FPR error on the held-out-from-fit subset, computed as predicted FPR minus empirical FPR divided by empirical FPR.
  }
  \label{fig:validation}
\end{figure}

We validate on the Credit Card Fraud Detection benchmark, which contains 284{,}807 European credit-card transactions and 492 fraud cases (0.172\% positive rate) \citep{dalpozzolo2015}.
We load the dataset with \texttt{sklearn.datasets.fetch\_openml(name="creditcard", version=1)}, using OpenML \texttt{data\_id} 1597.
A logistic-regression detector with standardized features is trained on a stratified 30\% split (85{,}442 rows; 148 positives).
The remaining 70\% is the detector holdout (199{,}365 rows; 344 positives).
A 30\% stratified slice of the holdout (59{,}809 rows: 59{,}706 benign and 103 positives) serves as the calibration-fit subset.
The complementary held-out-from-fit subset (139{,}556 rows: 139{,}315 benign and 241 positives) supplies the independent values in \cref{tab:eval} and the primary blue curves in \cref{fig:validation}.
The first two figure panels show the calibration-fit subset separately for comparison.
Logistic regression is chosen over a boosted-tree baseline because its linear margin resolves deep-tail FPR cleanly.
The sigmoid output of a boosted-tree classifier saturates around $10^{-3}$ FPR on this data and truncates the ROC tail; the calibration contract holds under either detector, but only the linear detector exercises the full anchor range.

\begin{table}[!htb]
  \centering
  \caption{Training and held-out FPR at calibrated score anchors.
  Training FPR uses the 59{,}706 calibration-fit benign rows; held-out FPR uses the 139{,}315 benign rows not used to fit calibration.
  Negative relative error means held-out FPR was above target.}
  \label{tab:eval}
  \begin{tabular}{rrrrrr}
    \toprule
    Calibrated score & Target FPR & Raw score & Training FPR & Held-out FPR & Rel. error \\
    \midrule
    0.10 & 10\% & 0.0008 & 9.999\% & 10.053\% & $-0.52\%$ \\
    0.30 & 1\% & 0.0036 & 0.9999\% & 1.023\% & $-2.24\%$ \\
    0.50 & 0.1\% & 0.0207 & 0.1005\% & 0.1005\% & $-0.49\%$ \\
    0.70 & 0.01\% & 0.9993 & 0.0100\% & 0.0093\% & $+7.17\%$ \\
    \bottomrule
  \end{tabular}
\end{table}

On the held-out-from-fit subset, the observed relative FPR error was at most 2.3\% from 10\% down to 0.1\% FPR and 7.2\% at 0.01\% FPR.
At the 0.01\% target, 13 of 139{,}315 held-out benign rows scored at or above the corresponding raw-score threshold.
This count gives an observed FPR of 0.0093\% and an exact 95\% Clopper--Pearson interval of 0.0050\%--0.0160\% \citep{clopper1934}.
The reported 7.2\% describes the point-estimate error for this split, whereas the confidence interval describes its sampling uncertainty.
\Cref{tab:eval} reports both the calibration-fit FPR and the held-out-from-fit FPR at each anchor from the latest run, and \cref{fig:validation} shows the corresponding held-out-from-fit behavior across the full curve.
Because its benign rows were never touched by the calibration fit, the held-out-from-fit subset measures generalization without the fit subset's near-target FPR-by-construction.
Below 0.01\% the 59{,}706-benign fit subset hits its binomial precision floor (\cref{tab:sample-sizes}), and the anchor at $10^{-5}$ FPR would require the calibration-fit subset to scale to about 1.5M benign samples for $\pm 50\%$ planning precision.

TPR at each fixed FPR measures detection quality, not calibration quality.
The detector catches 83\% of fraud at 0.1\% FPR, so the calibrated operating points land on real detections rather than an empty curve.
A detector whose TPR collapsed at low FPR would still calibrate to the same score axis and still fail the shipping bar.
That separation is deliberate: calibration keeps the FPR at each threshold consistent across deployments while detection quality is decided separately by comparing TPR at a fixed FPR across candidate models.

\section{Related work}
\label{sec:related-work}

This work composes established calibration and thresholding tools into a pipeline whose released score carries an FPR meaning across the whole deployment range.
Prior work supplies pieces of that construction, but not a compact score-to-FPR artifact with fixed log-FPR anchors.

\textbf{Probability calibration.}
Platt scaling fits a sigmoid to SVM scores so they can be read as posterior class probabilities \citep{platt2000}.
Zadrozny and Elkan use isotonic regression as a nonparametric alternative to sigmoid fitting, sorting labeled examples by score and using pool-adjacent-violators to learn a monotone step function from score to class probability \citep{zadrozny2002}.
Our shipped spline has the same inference shape, raw score in and calibrated value out, but the fitted quantity is the benign tail probability $q(\tau)$ from \cref{eq:true-deployment-fpr} rather than $\Pr(Y=1 \mid \mathrm{score})$.
This target change matters because precision moves with the positive-class rate, while FPR is defined by benign scores only.

\textbf{Outlier-score normalization.}
Merlion includes anomaly score calibration as a post-processing module to improve interpretability in a time-series anomaly-detection library \citep{bhatnagar2023}.
Kriegel et al. translate arbitrary outlier factors to $[0,1]$ values that can be compared across detectors and used in ensembles \citep{kriegel2011}.
These methods output calibrated or normalized anomaly scores.
Our calibrator fits the benign tail distribution directly and pins the released score to $\log_{10}(\mathrm{FPR})$ values in \cref{tab:anchors}.

\textbf{FPR and risk control.}
Neyman--Pearson classifiers choose a decision rule for a specified type-I error budget.
Scott and Nowak give one-sided learning-theoretic bounds of the form $R_0(\hat h)\le \alpha+\epsilon$ with high probability, while Tong et al. choose an order-statistic threshold so the population type-I error exceeds $\alpha$ with probability at most $\delta$ \citep{scott2005,tong2018}.
Learn Then Test calibrates predictive algorithms to satisfy finite-sample risk guarantees without model retraining and includes type-I outlier-error control among its examples \citep{angelopoulos2025}.
Finn and Johnson's early radar detector controls the threshold as a function of sampled clutter estimates, and radar textbooks treat constant-false-alarm-rate methods as threshold-detection tools \citep{finn1968,richards2014}.
Inductive conformal anomaly detection computes nonconformity p-values from a proper training set and calibration scores, then raises an alarm below a chosen significance level with a false-alarm guarantee that depends on the update mode and IID assumptions \citep{laxhammar2015}.
Bates et al. construct conformal p-values for outlier testing and give a uniform confidence bound for FPR as the raw-score threshold varies \citep{bates2023}.
These methods certify risk or test validity, but they do not ship a compact score transform with fixed log-FPR anchors.
We estimate FPR across the score range so one released score can support block, alert, review, and logging thresholds in the same policy.

\textbf{Empirical-CDF post-processing.}
\citet{dadalto2024} also use empirical CDFs of in-distribution detector scores, but their endpoint is a hypothesis test in which each detector score becomes a $p$-value under the in-distribution null and multiple detector $p$-values are combined with Fisher's statistic plus Brown's correction for correlated tests.
CADET calibrates reconstruction-error anomaly scores against sample hardness and then applies a binary decision rule at an approximately tuned false-positive level \citep{deng2022}.
Both papers calibrate anomaly or OOD scores, but their endpoint is a detector or test statistic rather than a fixed score contract for binary classifier releases.

\textbf{Plotting-position corrections.}
Plotting positions address the rank labels used inside our first spline.
The empirical label $k/n$ is the observed calibration-set count, but a continuous benign distribution places the $k$-th largest score at mean fresh-sample FPR $k/(n+1)$ \citep{nistOrderStatisticMeans}.
Filliben's plotting positions approximate the median rank location and, after flipping from lower-tail rank to upper-tail FPR, give median-centered FPR labels \citep{filliben1975,nistProbabilityPlot}.
These corrections are upstream knot-label choices for \cref{sec:bias}, not competing calibrators.

\textbf{Pipeline composition.}
The isotonic primitive, empirical-CDF transform, plotting-position correction, and monotone anchoring each predate this work.
The pipeline combines whole-curve FPR calibration on benign scores, a fixed piecewise-linear output contract in $\log_{10}(\mathrm{FPR})$ space, a deterministic artifact under 200 KB in our measurements, below-floor extrapolation, and release packaging that ships the calibrator alongside the model.

\section{Limitations}
\label{sec:limitations}

\emph{Score-granularity floor.}
A threshold cannot separate samples with the same score.
Here, 24 benign samples share the maximum score, so the detector jumps from zero FPR to $24/59{,}706 = 0.0402\%$ FPR.

\emph{Extrapolation below the sample-size floor.}
Calibration below the fit set's lowest observed tail FPR is linear extrapolation from the two lowest observed points.
For production FPR targets below the floor, fit calibration on a larger benign set rather than rely on extrapolation.

\emph{Benign distribution drift.}
The stability of the FPR contract assumes production benign traffic resembles calibration-time benign traffic.
Drift in legitimate user behavior shifts every threshold's actual FPR.
This is a data-freshness concern, not a defect of the calibrator, and it applies regardless of calibration method.

\emph{Plotting-position refinement.}
The implementation uses Filliben's median plotting position \citep{filliben1975} by default and exposes the mean-centered $k/(n+1)$ position as an option.
This adjustment fixes the FPR label assigned to rank-selected thresholds, but it does not reduce the finite-sample spread, so it does not change the sample-size requirements in \cref{tab:sample-sizes}.

\section{Usage}
\label{sec:usage}

Fitting, offline:

\begin{lstlisting}[language=Python]
from fpr_model_calibration.calibration import fit_calibration_pipeline
import joblib

pipeline = fit_calibration_pipeline(benign_scores, n_knots=10000)
joblib.dump(pipeline, 'calibration.pkl')
\end{lstlisting}

Inference, production:

\begin{lstlisting}[language=Python]
import joblib

pipeline = joblib.load('calibration.pkl')
calibrated = pipeline.predict(scores.reshape(-1, 1))
\end{lstlisting}

The log anchors ($0.5 = 0.1\%$ FPR, $0.7 = 0.01\%$, and so on) are tuned to AI security, where 1-in-1{,}000 is a practical planning point for automated action.
For domains with a different operational block threshold, edit \texttt{\_FPR\_CAL\_KNOTS} in \texttt{src/fpr\_model\_calibration/calibration.py} so that $0.5$ sits at the target FPR.

\section{Conclusion}
\label{sec:conclusion}

FPR calibration should give operators a stable alert-rate meaning for every score threshold, not only a calibrated probability at one operating point.
This method does that with benign-only data, a fixed log-scale FPR-to-calibrated-score contract, and two sklearn linear splines packaged as a standard \texttt{Pipeline}.
The first spline converts target FPRs to rescaled-score thresholds during fitting.
The second spline is the shipped raw-score-to-calibrated-score pipeline used at inference.

The evaluation shows that a fixed 10K-knot artifact preserves the calibrated curve when the benign calibration set supports the target FPR range, with the serialized artifact staying below 200 KB in our measurements.

The main constraint is benign sample count.
Filliben plotting positions place first-spline knots more accurately, but the dominant low-FPR error is still finite-sample uncertainty.
The 10K-knot representation makes deployment cheap; for production targets below the observed FPR floor, the practical answer is more benign calibration data, not a more elaborate spline.

\section*{Acknowledgments}
\label{sec:acknowledgments}

This work was prepared with AI assistance.
AI coding assistants and large language models were used for code review and refactoring of the accompanying software, prose editing of the manuscript, and citation verification.
The author reviewed and verified all technical content, results, and final wording, and bears full responsibility for the work.

\bibliographystyle{unsrtnat}
\bibliography{references}

@incollection{platt2000,
  author    = {Platt, John C.},
  title     = {Probabilities for {SV} Machines},
  booktitle = {Advances in Large-Margin Classifiers},
  editor    = {Smola, Alexander J. and Bartlett, Peter L. and Sch{\"o}lkopf, Bernhard and Schuurmans, Dale},
  publisher = {MIT Press},
  address   = {Cambridge, MA},
  pages     = {61--74},
  year      = {2000},
  isbn      = {9780262194488},
  doi       = {10.7551/mitpress/1113.003.0008},
}

@inproceedings{zadrozny2002,
  author    = {Zadrozny, Bianca and Elkan, Charles},
  title     = {Transforming Classifier Scores into Accurate Multiclass Probability Estimates},
  booktitle = {Proceedings of the 8th ACM SIGKDD International Conference on Knowledge Discovery and Data Mining (KDD)},
  publisher = {ACM},
  address   = {Edmonton, Alberta, Canada},
  pages     = {694--699},
  year      = {2002},
  doi       = {10.1145/775047.775151},
  isbn      = {158113567X},
}

@article{laxhammar2015,
  author    = {Laxhammar, Rikard and Falkman, G{\"o}ran},
  title     = {Inductive Conformal Anomaly Detection for Sequential Detection of Anomalous Sub-Trajectories},
  journal   = {Annals of Mathematics and Artificial Intelligence},
  volume    = {74},
  number    = {1--2},
  pages     = {67--94},
  year      = {2015},
  publisher = {Springer},
  doi       = {10.1007/s10472-013-9381-7},
  issn      = {1012-2443},
}

@article{bates2023,
  author    = {Bates, Stephen and Cand{\`e}s, Emmanuel and Lei, Lihua and Romano, Yaniv and Sesia, Matteo},
  title     = {Testing for Outliers with Conformal $p$-Values},
  journal   = {The Annals of Statistics},
  volume    = {51},
  number    = {1},
  pages     = {149--178},
  year      = {2023},
  month     = feb,
  publisher = {Institute of Mathematical Statistics},
  doi       = {10.1214/22-AOS2244},
  issn      = {0090-5364},
}

@article{angelopoulos2025,
  author    = {Angelopoulos, Anastasios N. and Bates, Stephen and Cand{\`e}s, Emmanuel J. and Jordan, Michael I. and Lei, Lihua},
  title     = {Learn Then Test: Calibrating Predictive Algorithms to Achieve Risk Control},
  journal   = {The Annals of Applied Statistics},
  volume    = {19},
  number    = {2},
  pages     = {1641--1662},
  year      = {2025},
  month     = jun,
  publisher = {Institute of Mathematical Statistics},
  doi       = {10.1214/24-AOAS1998},
  issn      = {1932-6157},
}

@article{scott2005,
  author    = {Scott, Clayton and Nowak, Robert},
  title     = {A {N}eyman--{P}earson Approach to Statistical Learning},
  journal   = {IEEE Transactions on Information Theory},
  volume    = {51},
  number    = {11},
  pages     = {3806--3819},
  year      = {2005},
  publisher = {IEEE},
  doi       = {10.1109/TIT.2005.856955},
  issn      = {0018-9448},
}

@article{tong2018,
  author    = {Tong, Xin and Feng, Yang and Li, Jingyi Jessica},
  title     = {{Neyman--Pearson} Classification Algorithms and {NP} Receiver Operating Characteristics},
  journal   = {Science Advances},
  volume    = {4},
  number    = {2},
  pages     = {eaao1659},
  year      = {2018},
  publisher = {American Association for the Advancement of Science},
  doi       = {10.1126/sciadv.aao1659},
  issn      = {2375-2548},
}

@article{finn1968,
  author    = {Finn, H. M. and Johnson, R. S.},
  title     = {Adaptive Detection Mode with Threshold Control as a Function of Spatially Sampled Clutter-Level Estimates},
  journal   = {RCA Review},
  volume    = {29},
  number    = {3},
  pages     = {414--464},
  month     = sep,
  year      = {1968},
  publisher = {RCA Laboratories},
}

@book{richards2014,
  author    = {Richards, Mark A.},
  title     = {Fundamentals of Radar Signal Processing},
  edition   = {2nd},
  publisher = {McGraw-Hill Education},
  address   = {New York, NY},
  year      = {2014},
  isbn      = {9780071798327},
}

@article{bhatnagar2023,
  author    = {Bhatnagar, Aadyot and Kassianik, Paul and Liu, Chenghao and Lan, Tian and Yang, Wenzhuo and Cassius, Rowan and Sahoo, Doyen and Arpit, Devansh and Subramanian, Sri and Woo, Gerald and Saha, Amrita and Jagota, Arun Kumar and Gopalakrishnan, Gokulakrishnan and Singh, Manpreet and Krithika, K. C. and Maddineni, Sukumar and Cho, Daeki and Zong, Bo and Zhou, Yingbo and Xiong, Caiming and Savarese, Silvio and Hoi, Steven and Wang, Huan},
  title     = {Merlion: End-to-End Machine Learning for Time Series},
  journal   = {Journal of Machine Learning Research},
  volume    = {24},
  number    = {226},
  pages     = {1--6},
  year      = {2023},
  publisher = {JMLR},
  issn      = {1533-7928},
  url       = {http://jmlr.org/papers/v24/22-0809.html},
}

@inproceedings{deng2022,
  author    = {Deng, Ailin and Goodge, Adam and Ang, Lang Yi and Hooi, Bryan},
  title     = {{CADET}: Calibrated Anomaly Detection for Mitigating Hardness Bias},
  booktitle = {Proceedings of the 31st International Joint Conference on Artificial Intelligence (IJCAI)},
  editor    = {De Raedt, Luc},
  publisher = {International Joint Conferences on Artificial Intelligence Organization},
  address   = {Vienna, Austria},
  pages     = {2002--2008},
  year      = {2022},
  doi       = {10.24963/ijcai.2022/278},
  isbn      = {978-1-956792-00-3},
}

@inproceedings{kriegel2011,
  author    = {Kriegel, Hans-Peter and Kr{\"o}ger, Peer and Schubert, Erich and Zimek, Arthur},
  title     = {Interpreting and Unifying Outlier Scores},
  booktitle = {Proceedings of the 11th SIAM International Conference on Data Mining (SDM)},
  publisher = {SIAM},
  address   = {Mesa, AZ, USA},
  pages     = {13--24},
  year      = {2011},
  doi       = {10.1137/1.9781611972818.2},
  isbn      = {9780898719925},
}

@article{dadalto2024,
  author    = {Dadalto C{\^a}mara Gomes, Eduardo and Alberge, Florence and Duhamel, Pierre and Piantanida, Pablo},
  title     = {Combine and Conquer: A Meta-Analysis on Data Shift and Out-of-Distribution Detection},
  journal   = {Transactions on Machine Learning Research},
  year      = {2024},
  publisher = {OpenReview},
  issn      = {2835-8856},
  url       = {https://openreview.net/forum?id=VGNBUS9TrU},
  doi       = {10.48550/arXiv.2406.16045},
}

@article{filliben1975,
  author    = {Filliben, James J.},
  title     = {The Probability Plot Correlation Coefficient Test for Normality},
  journal   = {Technometrics},
  volume    = {17},
  number    = {1},
  pages     = {111--117},
  year      = {1975},
  publisher = {Taylor \& Francis},
  doi       = {10.1080/00401706.1975.10489279},
  issn      = {0040-1706},
}

@misc{nistBinomialConfidence,
  author       = {{NIST/SEMATECH}},
  title        = {Confidence Intervals},
  howpublished = {e-Handbook of Statistical Methods, Section 7.2.4.1},
  year         = {n.d.},
  url          = {https://www.itl.nist.gov/div898/handbook/prc/section2/prc241.htm},
  note         = {Accessed 2026-05-28},
}

@misc{nistOrderStatisticMeans,
  author       = {{NIST/SEMATECH}},
  title        = {Order Statistics Means},
  howpublished = {Dataplot Reference Manual},
  year         = {n.d.},
  url          = {https://www.itl.nist.gov/div898/software/dataplot/refman2/auxillar/ordstmea.htm},
  note         = {Accessed 2026-05-07},
}

@misc{nistProbabilityPlot,
  author       = {{NIST/SEMATECH}},
  title        = {Probability Plot},
  howpublished = {e-Handbook of Statistical Methods},
  year         = {n.d.},
  url          = {https://www.itl.nist.gov/div898/handbook/eda/section3/probplot.htm},
  note         = {Accessed 2026-05-07},
}

@article{clopper1934,
  author    = {Clopper, C. J. and Pearson, E. S.},
  title     = {The Use of Confidence or Fiducial Limits Illustrated in the Case of the Binomial},
  journal   = {Biometrika},
  volume    = {26},
  number    = {4},
  pages     = {404--413},
  month     = dec,
  year      = {1934},
  publisher = {Oxford University Press on behalf of the Biometrika Trust},
  doi       = {10.1093/biomet/26.4.404},
  issn      = {0006-3444},
}

@inproceedings{dalpozzolo2015,
  author    = {Dal Pozzolo, Andrea and Caelen, Olivier and Johnson, Reid A. and Bontempi, Gianluca},
  title     = {Calibrating Probability with Undersampling for Unbalanced Classification},
  booktitle = {Proceedings of the IEEE Symposium Series on Computational Intelligence (SSCI)},
  publisher = {IEEE},
  address   = {Cape Town, South Africa},
  pages     = {159--166},
  year      = {2015},
  doi       = {10.1109/SSCI.2015.33},
  isbn      = {978-1-4799-7560-0},
}

@inproceedings{swanda2025,
  author    = {Swanda, Adam and Chang, Amy and Chen, Alexander and Burch, Fraser and Kassianik, Paul and Berlin, Konstantin},
  title     = {A Framework for Rapidly Developing and Deploying Protection Against Large Language Model Attacks},
  booktitle = {Proceedings of the 2025 Conference on Applied Machine Learning for Information Security (CAMLIS)},
  series    = {Proceedings of Machine Learning Research},
  volume    = {299},
  pages     = {200--221},
  year      = {2025},
  publisher = {PMLR},
  address   = {Arlington, VA, USA},
  issn      = {2640-3498},
  url       = {https://proceedings.mlr.press/v299/swanda25a.html},
  doi       = {10.48550/arXiv.2509.20639},
}

\appendix
\numberwithin{equation}{section}
\numberwithin{figure}{section}
\numberwithin{table}{section}

\section{Sample-size planning rule}
\label{sec:sample-size-rule-derivation}

The sample-size table uses the normal approximation in \cref{eq:normal-binomial-interval} as a planning rule.
Use the Clopper--Pearson interval \citep{clopper1934} for exact intervals.
At target FPR $p$, the 95\% half-width is approximately
\begin{equation}
\label{eq:appendix-absolute-half-width}
2\sqrt{\frac{p(1-p)}{n}}.
\end{equation}
In the low-FPR regime, $p(1-p)\approx p$, so the half-width becomes
\begin{equation}
\label{eq:appendix-tail-half-width}
2\sqrt{\frac{p}{n}}.
\end{equation}
A relative half-width of $r$ means that the absolute half-width is at most $rp$.
Setting \cref{eq:appendix-tail-half-width} equal to $rp$ and solving gives
\begin{equation}
\label{eq:appendix-sample-size-solve}
2\sqrt{\frac{p}{n}}=rp
\quad\Longrightarrow\quad
n=\frac{4}{r^2p}.
\end{equation}
For example, $p=10^{-3}$ and $r=0.5$ gives $n=16{,}000$ benign samples.

\section{Plotting-position simulation}
\label{sec:plotting-position-simulation}

\begin{figure}[!htb]
  \centering
  \includegraphics[width=0.95\textwidth]{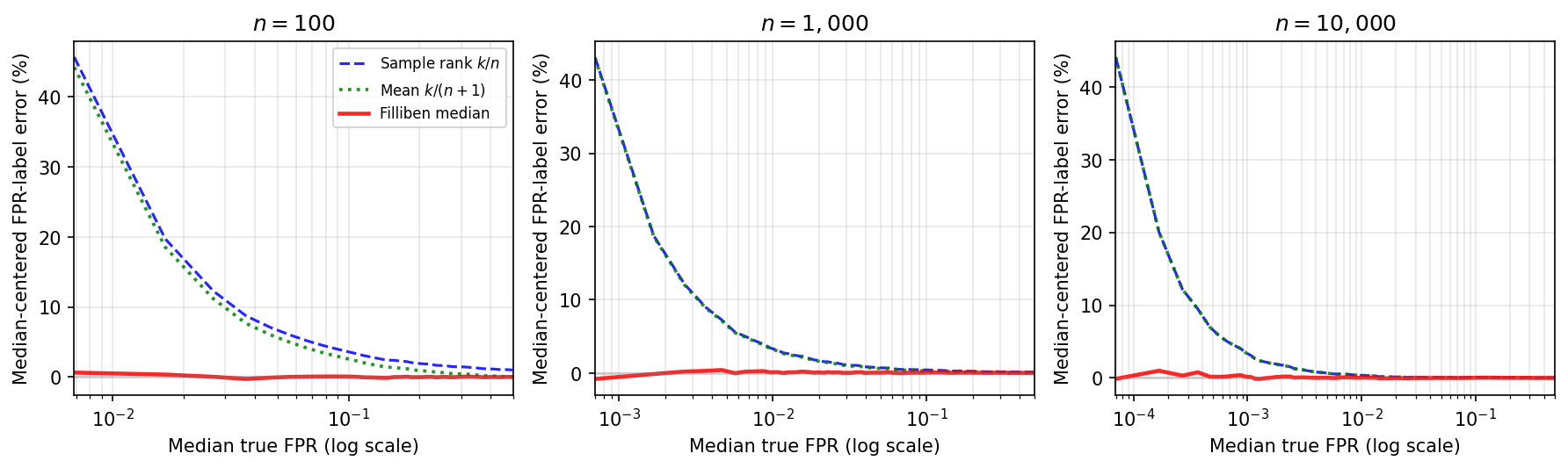}
  \caption{Median-centered FPR-label error for threshold $\tau_k$, the $k$-th largest benign model score.
  Each experiment draws $n$ benign scores from $\mathrm{Uniform}(0,1)$, selects the threshold $\tau_k$, and computes its fresh-sample FPR as $q_k=1-\tau_k$.
  Because the uniform distribution has survival function $\Pr(S\ge \tau)=1-\tau$, this measures $q_k$ directly rather than estimating it from another finite sample.
  For each plotting-position rule $m\in\{\mathrm{sample},\mathrm{mean},\mathrm{Filliben}\}$, the plotted error is $\widetilde q_k^{m}-\operatorname{median}(q_k)$, divided by $\operatorname{median}(q_k)$, across 50{,}000 repeated calibration samples.}
  \label{fig:plotting-position-error}
\end{figure}

\begin{figure}[!htb]
  \centering
  \includegraphics[width=0.95\textwidth]{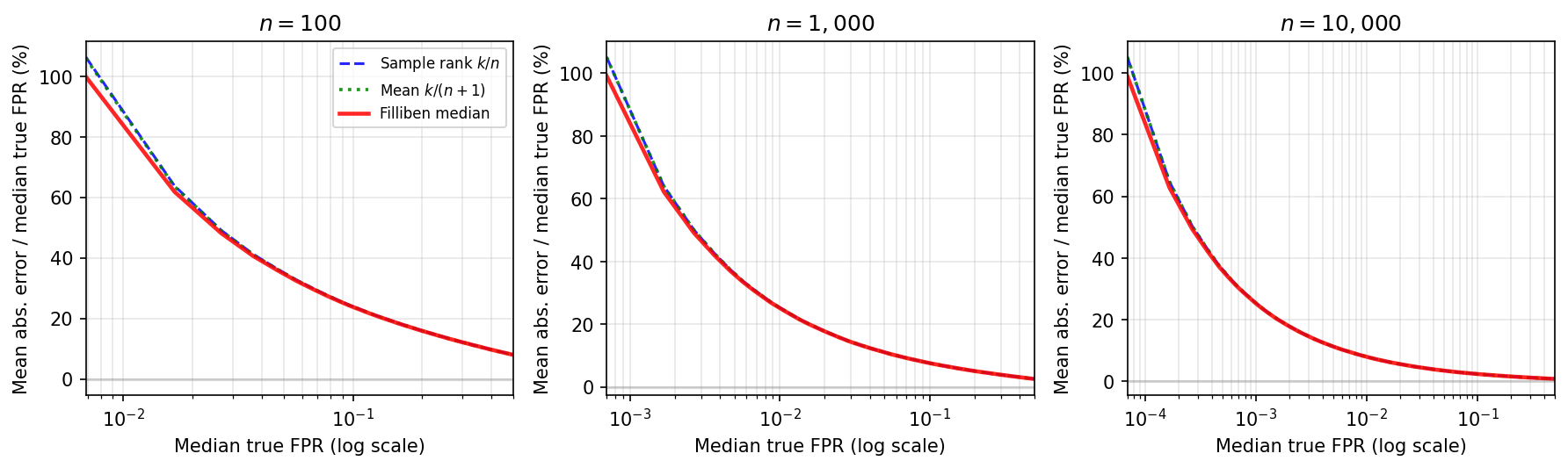}
  \caption{Mean absolute relative FPR-label error for the same simulation as \cref{fig:plotting-position-error}.
  For each plotting-position rule $m$ and rank $k$, the plotted value is the average of $|\widetilde q_k^{m}-q_k|$ across repeated calibration samples, divided by the median fresh-sample FPR for that rank.
  The denominator uses the median fresh-sample FPR rather than the per-draw fresh-sample FPR because rare maximum-score draws can make $q_k$ arbitrarily close to zero.
  This plot includes both the label-centering error and the finite-sample spread of the rank-selected threshold.}
  \label{fig:plotting-position-absolute-error}
\end{figure}

\end{document}